\documentclass{aa}  
\usepackage{graphicx}
\usepackage{natbib}
\bibpunct{(}{)}{;}{a}{}{,}  
\usepackage{txfonts}
\begin{document}
   \title{Envelope density pattern around wide binary AGB stars: A dynamical model}

   \author{J.H. He
          \inst{1}
          \inst{2}
          }

   \offprints{J.H. He}

   \institute{
              Academia Sinica, Institute of Astronomy and Astrophysics, 
              P.O. Box 23-141, Taipei 10617, Taiwan, China
              \thanks{\emph{This is my current address. \email{jhhe@asiaa.sinica.edu.tw}}}
         \and
              National Astronomical Observatories/Yunnan Observatory, Chinese Academy of Sciences, 
              P.O. Box 110, Kunming, Yunnan Province 650011, China
             }

   \date{Received September 19, 2006; accepted September 19, 2006}

 
  \abstract
   {Various morphologies such as multi-concentric shells and spiral-like patterns have been 
    observed around Proto-Planetary Nebulae and AGB stars. It is widely argued 
    that the regular density patterns are produced by binary systems.} 
   {The goal is to build up a simple dynamical model for the out-flowing 
    circumstellar envelope around AGB stars 
    in a wide binary system to explore the parameter dependence of the geometrical 
    characteristics of column density patterns.}
   {For an AGB star in a wide binary system, the orbital motion of the star can be 
    approximated as a series of pistons that simultaneously push out dust 
    and gas along radial directions, but work in different oscillation phases. 
    The piston model can fast produce column density patterns with high enough spatial 
    resolutions for parameter dependence exploration.}
   {The formation of 3-D quasi-spherical density structure is induced by orbital motion of the AGB star.
    The column density pattern only depends on two parameters: eccentricity 
    of the orbit $e$ and the terminal outflow velocity to mean orbital velocity ratio $\gamma$. 
    When viewed perpendicular to the orbital plane, \emph{spiral}, \emph{broken spiral}, and 
    \emph{incomplete concentric shell patterns} can be seen, while when viewed along the orbital 
    plane, \emph{alternative concentric half-shell}, \emph{egg-shell}, and \emph{half-shell half-gap 
    patterns} will develop. Non-zero eccentricity causes asymmetry, while larger $\gamma$ makes a 
    weaker pattern and helps bring out asymmetry. A spiral pattern may becomes broken when $e>0.4$. The 
    spiral center is always less than $12\%$ of spiral pitch away from the orbit center.
    One should have more chances ($\sim 80\%$) seeing spiral-like patterns than seeing concentric 
    shells ($\sim 20\%$) in the circumstellar envelope of wide binary AGB stars.}
   {}


   \maketitle
%

\section{Introduction}
\label{intro}

In recent years, various kinds of morphology in circumstellar envelopes of AGB stars have been 
revealed by high spatial resolution observations. 
For example, \citet*{mauron00} showed a series of shells in the envelope of the 
carbon rich AGB star \object{IRC +10216} in an optical image. 
\citet{balick} showed clear multi-shell structures in the relic AGB envelope around 
the planetary nebula \object{NGC 6543}, the Cat's Eye. 
\citet{kwok} found multi-arcs near the two lobes of a bipolar proto-planetary nebula 
\object{IRAS 17150-3224}. High velocity bipolar jets 
had also been discovered in the AGB circumstellar envelope by means of interferometry of maser 
emission, e.g., the observations by \citep{imai02,imai05} towards \object{W 43A}. 
The most marvelous pattern found most recently is the spiral 
pattern found by \citet*{mauron06} in the envelope of \object{AFGL 3068}, a C-rich AGB star. 
They fitted an Archimedes' spiral to the observed spiral pattern, which demonstrated the 
amazing regularity of the pattern. 

The mechanism that produces the various morphologies is still controversial. \citet*{steffen} 
proposed periodic change of mass-loss rate or wind speed, but the change of mass-loss 
rate by thermal pulse has too long a period ($\sim 10^5$ yrs). 
\citet{balick} attributed the multi-shells around the 
Cat's Eye to instability of physical and chemical processes during dust 
formation and growth, as suggested by \citet{simis}, who tried to explain the concentric 
shells around \object{IRC +10216} in the same way. \citet*{mauron06} preferred to explain their spiral 
pattern in \object{AFGL 3068} by numerical results from \citet*{mastrodemos}, (MM99) hereafter, in which 
an AGB star in a wide binary produces a spiral-like density pattern by the 
reflex motion of the star (their model 4). MM99 performed smooth particle hydrodynamic (SPH) 
simulations to explore the effects of binarity on AGB wind for circular orbit binaries. 
In their wide binary example, they showed spiral and concentric shell density patterns formed 
in the circumstellar envelope. But to understand the observed patterns better, it's 
helpful to know, if the orbit is not circular, what will be different in the patterns. However, 
due to difficulties in SPH simulation, e.g., time consuming, not high enough spatial 
resolution, varying of the particle's smoothing length, etc., a simpler simulation is needed for 
extensive parameter exploration tasks.

\Citet{soker} discussed the influence of a wide binary on the structures of 
planetary nebulae by taking the AGB mass-loss as a sequence of mass loss pulses. With 
far distance piston approximation and sticky particle model in my work, 
the influence of binarity will be explored
in more detail. Section~\ref{model} gives the numerical piston model to simulate the 
formation of density pattern 
in the circumstellar envelope of a wide binary AGB star. Numerical 
results are present in Sect.~\ref{results}.  A brief summary is given in 
Sect.~\ref{summary}.

\section{Piston model and column density pattern}
\label{model}

\subsection{Piston model}
\label{piston}

To simplify the problem, several assumptions about an AGB star are made: mass-loss rate does 
not change with time; outflow velocity is constant and equal to terminal velocity $V_\mathrm{e}$; 
gas pressure can be neglected; and dust gas interaction can be neglected. Although the mass-loss 
rate of AGB stars may change with time due to thermal pulse, it changes on a very 
long timescale \citep*[e.g., $10^5$ years, as suggested by][]{steffen} that is much longer 
than the orbital period. Mass-loss rate and outflow velocity may change due to the interplay of 
dust formation and dynamics \citep{simis}, but it is not clear whether such a phenomenon really 
occurs. The assumption of constant outflow velocity is equivalent to assuming that the gravity 
force and radiation pressure on dust grains and gas are balanced everywhere. This is supported 
by hydrodynamic calculations \citep[e.g., the model results for \object{IRC +10216} by][]{skinner}, 
unless one examines the region very close the star. Gas pressure can broaden any density 
structure by means of thermal dissipation, but its effect is marginal in the outer part of the supersonic 
outflow. And gas pressure does not alter the geometry of density structures, 
which is the main topic of this work. Gas and dust are not distinguished and the material is 
always called gas. These assumptions sound plausible for this work because hydrodynamic 
effects on the pattern geometry become unimportant in wide binary systems.

Under these assumptions, the motion of each chunk of ejected gas can be treated as ballistic motion. 
One can imagine that many gas parcels are ejected from the star surface with constant 
velocity $V_\mathrm{e}$ relative to the star. The absolute velocity of each gas parcel in the 
center of mass (CoM) frame of the binary is equal to the 
sum of $V_\mathrm{e}$ and star motion velocity.

This work only considers density patterns formed far away from the star in the circumstellar 
envelope. In this case, the star motion and mass loss can be approximated as a piston that is 
ejecting gas parcels and moving back and forth along radial direction to modulate the velocity 
distribution in the ejected gas. The ejected gas will be compressed by the velocity gradient to 
form a density enhancement. Shock may happen when gas parcels collide with each other due to 
different velocities. One such piston works in one direction. The whole three-dimensional AGB 
mass-loss envelope can be imagined as a collection of pistons pointing in different directions 
and running in different phases. The different piston phases in different directions concatenate 
the density enhancements or shocked regions to form a three-dimensional quasi-spherical density 
structure, like the structure produced by the SPH simulations by 
MM99 for circular orbit in their model 4. This is called a 
{\it far distance piston approximation}.
   \begin{figure}
   \centering
   \begin{minipage}[c]{2in}
      \centering
      \includegraphics[bb=13 38 214 196, width=2in,clip]{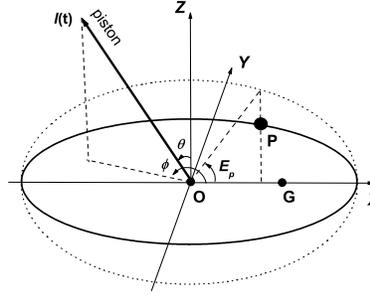}
   \end{minipage}
   \begin{minipage}[c]{1.5in}
      \caption{Schematic illustration of the orbital motion of AGB star (point P), 
               with definitions of Cartesian coordinates OXYZ, eccentric anomaly $E_{\mathrm{p}}$, 
               and position angle $(\theta,\phi)$ of the piston (the heavy arrow). Point O is the 
               center of the orbit while G is the center 
               of mass. The dotted circle is the circumscribed circle of the orbit.}
      \label{figorbit}
   \end{minipage}
   \end{figure}

The orbital motion of the AGB star is assumed to be Keplerian. Projection of the star motion 
along any given direction $(\theta,\phi)$ (see definition of quantities in Fig.~\ref{figorbit}) 
gives the position of the piston along that direction as
\begin{equation}
\label{eqlt}
l(t) = a_{\mathrm{p}}\left(\cos{E_{\mathrm{p}}}\cos{\phi}+\sqrt{1-e^2}\sin{E_{\mathrm{p}}}\sin{\phi}\right)\sin{\theta}.
\end{equation}
Here $e$ and  $a_p$ are the eccentricity and semi-major axis of the star orbit, respectively, while 
$E_p$ is the eccentric anomaly of the star. Then the speed of the piston is 
\begin{equation}
\label{eqvlt}
V_{l}(t) =\frac{\mathrm{d}l(t)}{\mathrm{d}t}
         =\omega a_{\mathrm{p}}\frac{\sqrt{1-e^2}\cos{E_{\mathrm{p}}}\sin{\phi}-\sin{E_{\mathrm{p}}}\cos{\phi}}{1-e\cos{E_{\mathrm{p}}}}\sin{\theta},
\end{equation}
where $\omega=2\pi/T$ is the mean angular velocity of the star motion with orbital period $T$. 
The time dependence of piston motion is reflected by
\begin{equation}
\label{eqept}
E_{\mathrm{p}}-e\sin{E_{\mathrm{p}}}=\omega t.
\end{equation}

In the numerical simulation, the $4\pi$ solid angle is divided into 10000 equally sized patches 
and a piston is computed along the center direction of each solid angle element. Each 
piston is assumed to eject 100 gas parcels (with equal mass) within one orbital period. 
Gas parcels ejected at different times set out from different positions $l(t)$ with different 
velocities $V_{\mathrm{e}}+V_{l}(t)$. After leaving the star, each gas parcel moves outward 
with constant speed, until it collides with neighboring parcels. The collision of gas parcels 
causes shock and the collided gas parcels are arbitrarily assumed to merge with each other 
(actually move together side by side with very small fixed distances in
the simulation, sticky particles). 
The velocity of the merged gas parcels is determined by conservation of momentum. The spatial 
distribution of all gas parcels represents the 3D density distribution of out-flowing material 
in the AGB envelope. Projection of the 3D distribution of gas parcels onto any specified 
sky plane produces a column density pattern.

\subsection{Normalization of the problem}
\label{normalization}

Several physical parameters are involved in the piston model: orbital parameters 
such as semi-major axis $a_{\mathrm{p}}$, eccentricity $e$, and period $T$, and the steady 
outflow velocity $V_{\mathrm{e}}$. However, not all of them are independent of the problem. 
To reduce the number of parameters, the above formulae need be normalized by dividing 
all length quantities by the circumference of the circumscribed circle of the orbit 
$2\pi a_{\mathrm{p}}$, dividing all time quantities by period $T$, and dividing all velocity 
quantities by the mean orbital velocity of the star $\omega a_{\mathrm{p}}$. Then, the 
piston motion formulae Eqs.~(\ref{eqlt}) and (\ref{eqvlt}) become
\begin{equation}
\label{eqltn}
\tilde{l}(\tilde{t}\,) =\frac{1}{2\pi}\left(\cos{E_{\mathrm{p}}}\cos{\phi}+\sqrt{1-e^2}\sin{E_{\mathrm{p}}}\sin{\phi}\right)\sin{\theta}
\end{equation}
\begin{equation}
\label{eqvltn}
\tilde{V}_{l}(\tilde{t}\,) =\frac{\sqrt{1-e^2}\cos{E_{\mathrm{p}}}\sin{\phi}-\sin{E_{\mathrm{p}}}\cos{\phi}}{1-e\cos{E_{\mathrm{p}}}}\sin{\theta}
\end{equation}
and
\begin{equation}
\label{eqt0n}
E_{p}-e\sin{E_{p}}=2\pi\tilde{t}.
\end{equation}
In the numerical simulation, each gas parcel is ejected at normalized position $\tilde{l}(\tilde{t})$ 
with the constant normalized speed $\gamma+\tilde{V}_l(\tilde{t})$ 
(with $\gamma=V_\mathrm{e}/\omega a_\mathrm{p}$ 
being a velocity ratio). After the normalization, 
one has only two free model parameters: $e$ and $\gamma$. All other parameters 
such as $a_{\mathrm{p}}$, $T$, and $V_\mathrm{e}$ are combined in the parameter $\gamma$.

\subsection{Formulation of single piston density pattern}
\label{pistond}

Thanks to the simplicity of the piston model, it is possible to directly 
formulate the density pattern geometry (position of density peaks or shocked regions) 
in the outflowing gas. In some cases, solving 
the formulas may be easier than analyzing numerical column density maps in the discussion of certain 
aspects of the pattern geometry. But due to the limitation of space, the deduction of normalized 
pattern formulas is put in Appendix~\ref{formulae} that is available only in electronic version.

\section{Results}
\label{results}

\subsection{Numerical column density patterns}
\label{cdmaps}

\begin{figure*}
   \begin{minipage}[c]{5.7in}
      \centering
      \includegraphics[bb=66 504 487 728, width=5.7in, clip]{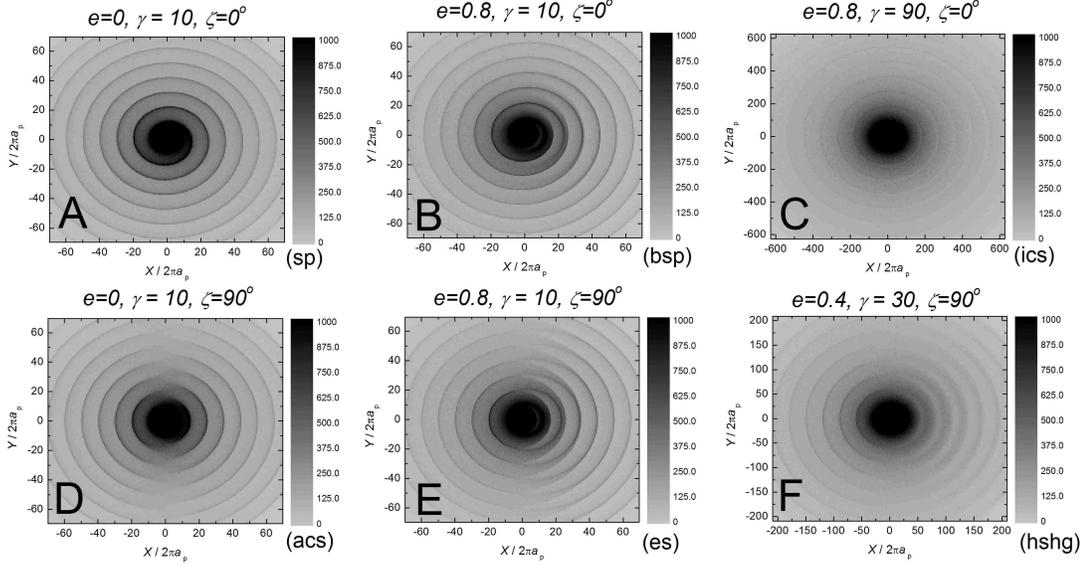}
   \end{minipage}
   \begin{minipage}[c]{1.5in}
      \centering
      \caption{Column density patterns calculated by the piston model for representative cases. 
            Observer's azimuth angle is always $\theta=-90\degr$, which means the pericenter 
            of star orbit is pointed to the right. Inclination angle $\zeta$ and $e, \gamma$ 
            are shown on top of each map. All maps are in linear 
            stretch in the same gray scale range. X and Y axes are in normalized length 
            unit (i.e., times of $2\pi a_p$). Pattern types shown at the right bottom of 
            each map are: \emph{sp} -- spiral; \emph{bsp} -- broken spiral; 
            \emph{ics} -- incomplete circular shells; \emph{acs} -- alternative concentric shells; 
            \emph{es} -- egg shells; and \emph{hshg} -- half-shell half-gap.}
   \label{mape}
   \end{minipage}
   \end{figure*}

Column density pattern geometry has been explored in a large range of parameter space: 
$e = 0.0-0.8$, $\gamma = 10-90$, and different observers' direction angles 
$(\zeta,\psi)$. Representative types of morphology found in the computed column density 
maps are shown in Fig.~\ref{mape}. In the first row patterns are viewed perpendicular 
to the orbital plane, while in the second row patterns are viewed along
the orbital 
plane. Maps A and D show that, for the circular orbit case, the column density pattern 
is a good \emph{spiral-like} pattern when viewed perpendicular to the orbital plane 
and \emph{alternative concentric half shell} pattern when viewed along the orbital plane, 
which closely resembles the SPH simulation results of 
MM99 (their Model 4). Comparing map A with B tells us that, when 
eccentricity is non-zero, the spiral pattern may become broken in 
its inner part (the \emph{broken spiral pattern} in map B). The extent of the spiral break 
decreases when the shock pattern moves outwards, and eventually continuous spiral forms appear in 
the outer part of the pattern.
If both $e$ and $\gamma$ are large, as the case of map C, even if observed perpendicular to 
the orbit plane, the column density pattern still looks like a series of concentric shells, but 
the shells are not only much fainter, but also incomplete in the lower-right quarter 
(\emph{incomplete concentric shell pattern}). 

A common character of the patterns in the second row of
  Fig.~\ref{mape} is concentric shells with the 
lack of density pattern along the vertical directions, which are the directions of the orbit pole. 
Comparing maps D and E tells us that non-zero 
orbital eccentricity may cause the variation of relative radii of the concentric half-shells 
and make strong non-shocked patterns around shocked patterns. As a result, 
in map E, 
the radii of the left and right series of half-shells are nearly the same, which 
imitate nested \emph{egg shells}. In the right half of map E, the innermost shells look thicker. 
They are actually double shells with non-shocked gas filled in between. Moving outwards, 
the double shells eventually merge with each other to form sharp single-layer shocked shells. 
One should also have noticed the left-right asymmetry both in  
density contrast and in pattern shape in map E, which is the effect of non-zero eccentricity, too.
Comparing map F with maps D and E gives us a surprise: the larger $e$ and $\gamma$ 
in map F cause the missing of shells on the right side, which results in 
a \emph{half-shell half-gap} pattern. 

Exploration of more maps shows that left-right 
asymmetry (as in the broken spiral, incomplete concentric shell, egg shell, half-shell half-gap patterns) 
only appears when $e>0$ and $\gamma$ is large, otherwise, when $\gamma$ is small, even if $e>0$ 
and the asymmetry exists, the asymmetry may be buried in the invisible central region of the maps. 
When the asymmetry is prominent, 
its dependence on the observer's azimuth angle is also found when observing along 
the orbital plane: the pattern looks more left-right symmetrical when being viewed along the 
major orbit axis, while it appears more left-right asymmetrical when being viewed
along the minor orbit axis.
Several maps that illustrate this effect are present in Fig.~\ref{mapaz} in the electronic 
version only.
\onlfig{3}{
   \begin{figure}
      \centering
      \begin{minipage}[b]{1.9in}
      \centering
      \includegraphics[bb=159 299 322 735, width=1.7in,clip]{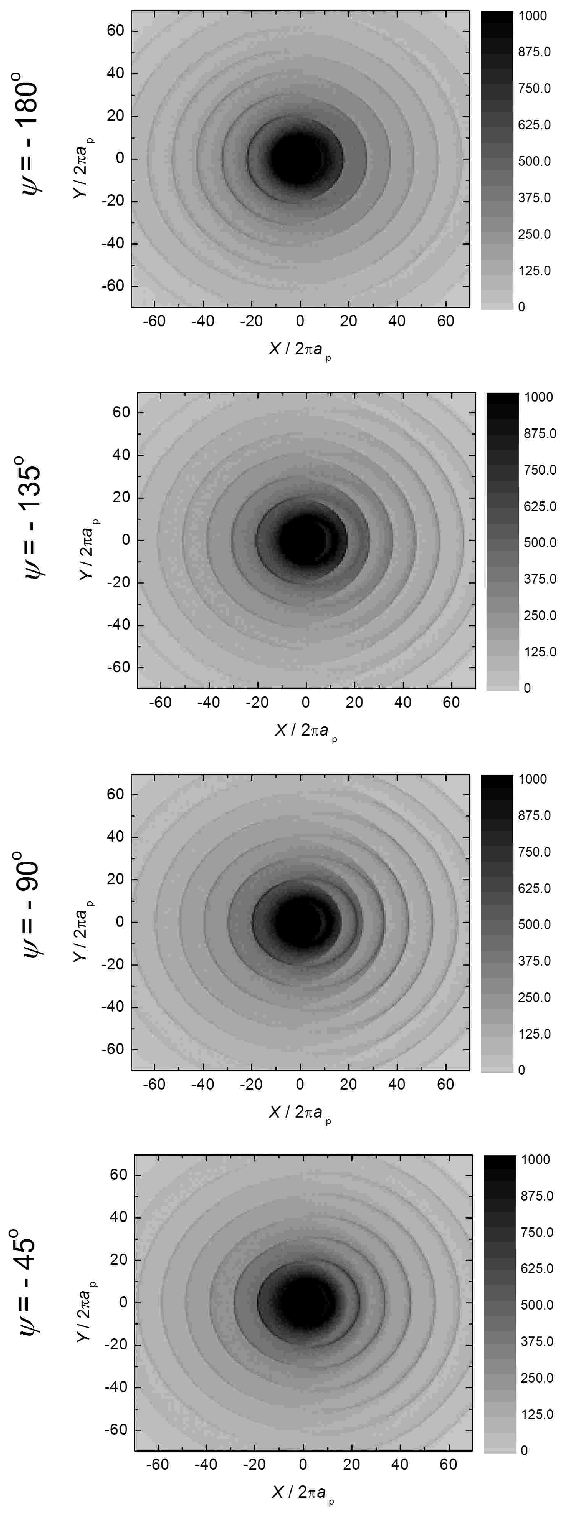}
      \end{minipage}
      \begin{minipage}[b]{1.3in}
      \centering
      \caption{Computed maps show the azimuth angle effects on pattern
        geometry. All maps are shown for $e=0.8, \gamma=10, \zeta=-90\degr$, 
        that is, viewed along the orbit plane (edge on). $\psi=-180\degr$ means viewing 
        along the major orbit axis, while $\psi=-90\degr$ means viewing along the 
        minor orbit axis.}
      \end{minipage}
   \label{mapaz}
   \end{figure}
}

\subsection{Several interesting properties of the column density patterns.}
\label{properties}

\begin{figure*}
   \centering
   \begin{minipage}[c]{5.7in}
      \includegraphics[bb=72 634 483 722, width=5.7in, clip]{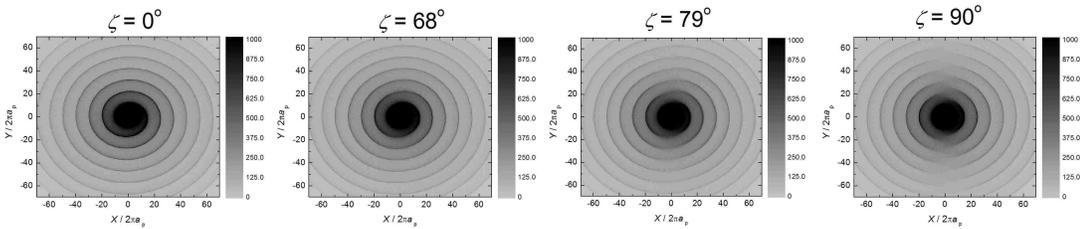}
   \end{minipage}
   \begin{minipage}[c]{1.5in}
      \caption{Maps show inclination angle effects. Azimuth angle is
            $\psi=-90\degr$, $e=0$, and $\gamma=10$ for all maps. One can see 
            the spiral pattern in the inclination angle range of $0\degr-79\degr$ and 
            the concentric shells in the $79\degr-90\degr$ range.}
      \label{figincl}
   \end{minipage}
   \end{figure*}

One of the interesting problems is how the transition from spiral-like pattern 
to concentric half shell pattern happens when the observer's inclination angle 
changes from $0\degr$ to $90\degr$. In Fig.~\ref{figincl}, more column density 
patterns are shown for different observation inclination angles 
$\zeta = 0\degr, 68\degr, 79\degr,$ and $90\degr$ for the circular orbit case. 
One can see that only when the inclination angle is $\zeta \geq 79\degr$, 
does the column density pattern begin to prominently deviate from the spiral-like pattern. 
That is to say, in a large range of inclination angles, from $0\degr$  
to $79\degr$, one can see a spiral-like pattern. If one sees a concentric shell 
pattern, the inclination angle of observation must be within the $90\pm10$ degree range, 
that is, the star orbit must be almost edge on. If the distribution of the star orbit orientation 
is completely random, one has an approximate 
probability of $1-\cos{79\degr} = 81\%$ seeing the spiral-like pattern and only a 
probability of $19\%$ seeing the concentric shell pattern in the circumstellar envelope 
of such a wide binary AGB star. MM99 intuitively speculated that,
when viewed with finite resolution at great distances from the star and 
from some intermediate latitude, the multi-shells may appear to form full circles. 
This is contradicted with the conclusion of this work and is incorrect. Actually
the relative shift
of shells with inclination angle is true but the shift of shells 
forms a spiral pattern, instead of
full rings. This can be easily understood because one sees spirals when observing perpendicular 
to the orbit plane and because the left and right shells have different radii.

\begin{figure}
   \centering
   \begin{minipage}[c]{2in}
      \includegraphics[bb=16 18 283 216, width=2in, clip]{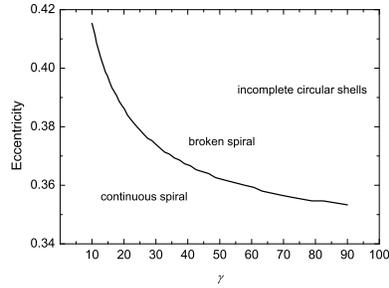}
   \end{minipage}
   \begin{minipage}[c]{1.5in}
      \caption{The border line divides the $\gamma-e$ parameter space into two parts: the density 
               pattern in the orbital plane of the model is a continuous spiral pattern 
               in the lowerleft part and a broken spiral or incomplete concentric shell 
               pattern in the upperright part.}
       \label{figbrokensp}
   \end{minipage}
   \end{figure}

Another question is when the spiral pattern becomes a broken spiral. The answer 
lies in the single piston model along the pericenter direction of the orbit
because the broken part of a spiral always happens on the pericenter side. Some 
trial solutions of single piston formulas~(\ref{eqsolvtp1n}, \ref{eqsolvtp2n}, \ref{eqsolvtp3n}) 
demonstrate that the formation of the broken spirals is due to a 
double solution of $E_p$ within one period. These formulas are numerically solved
for all possible combinations of $(e,\gamma)$. Those pairs of
$(e,\gamma)$ 
beyond which a double solution of $E_p$ begins to appear are shown in 
Fig.~\ref{figbrokensp} as a border line that divides the parameter
space into two halves: lower left half for the continuous spiral; upper
right half for the broken spiral and the incomplete concentric shell
pattern.
Because $e$ does not change too much around $0.4$ along the border line, one may 
say that a broken spiral pattern may appear when eccentricity $e \geq 0.4$.

The third question is where the spiral center is. Archimedes' spiral
curve is fitted to the mature part of shocked patterns computed
from the piston model in a wide range of parameter space
$(e=0-0.8,\gamma=2-90)$. 
The pattern fitting demonstrates that the shocked patterns are
always good Archimedes' spiral curves. 
The center of the spiral is not at the orbit
center, but the shift of the spiral center from the orbit center 
shows some degree of dependence on both parameters $e$ and $\gamma$. 
Generally speaking, higher eccentricity 
and smaller $\gamma$ causes larger spiral center shifts. However, in most cases, 
the spiral center shift is
always less than $2\%$ of a spiral pitch in the major orbit axis direction
and always less than $12\%$ of a spiral pitch in the minor orbit axis
direction, which means the spiral center should be just on the star
position when the observation resolution is not extremely high. In circular 
orbit cases, the spiral center is exactly at the orbit center.

The properties of the various column density patterns may be
  critical in distinguishing the binary mechanism
from the mass-loss rate modulation mechanism. The patterns 
also serve as a magnifier to show the binary motion in the
center of the circumstellar envelope. On the other hand, the formation of the density
pattern, especially the shocked density pattern, greatly enhances the
visibility of the circumstellar material and enables us to study
the mass-loss history of AGB stars in binary systems.

\section{Summary}
\label{summary}

A piston model is used to simulate the quasi-spherical density patterns formed 
in the supersonic outflow of wide binary AGB stars. The simple and quick model allows 
parameter space exploration in detail. Normalization of the piston model shows 
that the normalized 3-D density pattern in the circumstellar envelope only 
depends on two parameters: eccentricity of the star orbit $e$ and steady outflow 
velocity to mean star motion velocity ratio $\gamma$ ($=V_{\mathrm{e}}/\omega a_\mathrm{p}$). 
When viewed perpendicular to the orbital plane, the 
\emph{spiral-like pattern, broken spiral pattern}, and \emph{incomplete concentric shell pattern} 
can be seen, while when viewed along the orbital plane, the 
\emph{alternative concentric half-shell pattern, egg-shell pattern}, and \emph{half-shell half-gap pattern} 
can be seen. Higher eccentricity causes more {left-right asymmetry and
  the variation of relative 
radii between the left and right series of concentric half shells}, while larger $\gamma$ 
makes a weaker density pattern but helps bring out asymmetry from the pattern center.

One sees a spiral-like pattern in a large observers' 
inclination angle range $(0\degr-79\degr)$, while concentric half-shell patterns 
can be seen only in edge-on binary, which indicates that one has more of a 
chance ($\sim 81\%$) seeing spiral-like patterns in the sky than 
seeing concentric shells ($\sim 19\%$ chance). Fitting Archimedes' spiral curve 
to numerical shocked column density patterns demonstrates that the spiral center is away 
from the orbit center by no more than $12\%$ of the spiral pitch. The spiral pattern 
may become broken when the eccentricity is larger than about $0.4$.

\begin{acknowledgements}
       I thank the anonymous referee who helped me to condense and improve the descriptions 
       in this work. This work is partly supported by the National Natural Science Foundation of 
       China under Grant No. 10433030 and No. 10503011.
\end{acknowledgements}

\bibliographystyle{aa}  
\bibliography{6435}   

\Online

\begin{appendix}
\section{Formulation of single-piston density-pattern}
\label{formulae}

This section only considers the normalized piston model in which the normalized 
orbital period is $1$ and the normalized steady outflow velocity is $\gamma$. 
Pattern geometry is determined by the positions of all density peaks and/or shock regions.
Because the piston motion is periodic, one only needs to consider gas parcels ejected within 
a single period. 
Before the occurrence of shock, there should be no collision of gas parcels. One may find a 
\emph{peak parcel} that corresponds to a density maximum. Assuming the peak parcel was ejected at 
an earlier time $\tilde{t}_\mathrm{p}$ (the \emph{peak parcel ejection
  time}), the radial position of the peak parcel 
(or the density peak) at a given time $\tilde{t}$ is 
\begin{equation}
\label{eqrpeakn}
\tilde{r}_{\mathrm{peak}}(\tilde{t})=\tilde{l}(\tilde{t}_{\mathrm{p}})+\left[\gamma+\tilde{V}_{l}(\tilde{t}_{\mathrm{p}})\right](\tilde{t}-\tilde{t}_{\mathrm{p}}).
\end{equation}
Here the gas parcel velocity is expressed as two components: steady outflow velocity $\gamma$ and 
an individual velocity component
$\tilde{V}_{l}(\tilde{t}_{\mathrm{p}})$ contributed by the star motion. 
At a later time $\tilde{t}_{\mathrm{c}}$ when the first collision between gas parcels occurs 
(the peak parcel 
is usually the parcel that collides with others first), shock region 
begins to form. The time $\tilde{t}_{\mathrm{c}}$ is called the \emph{shock time}, and the parcels that 
collide first are called the ``\emph{shock parcels}". After that, further gas parcel 
collisions modify the velocity of the shock parcels according to the conservation of momentum. 
At any given time $\tilde{t} > \tilde{t}_{\mathrm{c}}$, assuming the shock parcels were 
ejected at an earlier time 
$\tilde{t}_{\mathrm{s}}$ (the \emph{shock parcel ejection time}), the
radial position of shock 
parcels (or shock region) is
\begin{equation}
\label{eqrshockn}
\tilde{r}_{\mathrm{shock}}(\tilde{t})=\tilde{l}(\tilde{t}_{\mathrm{s}})+\gamma(\tilde{t}-\tilde{t}_{\mathrm{s}})+\tilde{V}_{l}(\tilde{t}_{\mathrm{s}})(\tilde{t}_{\mathrm{c}}-\tilde{t}_{\mathrm{s}})+\int_{\tilde{t}_{\mathrm{c}}}^{\tilde{t}}{\tilde{V}_{\mathrm{m}}(\tilde{t})\mathrm{d}\tilde{t}}.
\end{equation}
The individual velocity component of the shock parcels $\tilde{V}_{\mathrm{m}}(\tilde{t})$ 
in the last term on the right side 
is different from star motion velocity $\tilde{V}_{l}(\tilde{t}_{\mathrm{s}})$ because collisions 
modify it according to the convervation of momentum. 
$\tilde{V}_{\mathrm{m}}(\tilde{t})$ cannot be analytically expressed, but 
has to be determined by piston model simulation.

To calculate the radial positions of density peaks or shock regions using the above formulas, one needs to  
know $\tilde{t}_{\mathrm{p}}, \tilde{t}_{\mathrm{s}}, \tilde{t}_{\mathrm{c}}$. They can be determined using 
the definition of density maximum. Under the assumption of constant mass-loss rate $\dot{M}$ and 
outflow velocity $\gamma$, the overall density distribution $\tilde{\rho}(\tilde{t},\tilde{t}_\mathrm{p})$ 
in the circumstellar envelope at a given time $\tilde{t}<\tilde{t}_\mathrm{c}$ 
roughly follows an inverse square law of radius $\tilde{r}(\tilde{t}_\mathrm{p})$ 
upon which the density patterns are superimposed. 
Therefore, it is helpful to define a pattern density by multiplying the square of radius by 
density. It is not difficult to deduce the formula of pattern density as 
\begin{equation}
\label{eqrhop2}
\tilde{\rho}^{\prime}(\tilde{t},\tilde{t}_\mathrm{p})=
\tilde{\rho}(\tilde{t},\tilde{t}_\mathrm{p})\tilde{r}^2(\tilde{t}_{\mathrm{p}})=
\frac{\dot{M}}{4\pi}\left/\left[\gamma-\frac{\mathrm{d}\tilde{V}_{l}(\tilde{t}_\mathrm{p})}{\mathrm{d}\tilde{t}_\mathrm{p}}(\tilde{t}-\tilde{t}_\mathrm{p})\right]\right..
\end{equation}
By constraining the maximum of the pattern density $\tilde{\rho}^{\prime}(\tilde{t},\tilde{t}_\mathrm{p})$, 
one may derive the following formulas for solving $\tilde{t}_\mathrm{p}$:
\begin{equation}
\label{eqsolvtp1n}
\frac{\mathrm{d}^{2}\tilde{V}_{l}(\tilde{t}_{\mathrm{p}})}{\mathrm{d}\tilde{t}_{\mathrm{p}}^{\,2}}\left(\tilde{t}-\tilde{t}_{\mathrm{p}}\right)=\frac{\mathrm{d}\tilde{V}_{l}(\tilde{t}_{\mathrm{p}})}{\mathrm{d}\tilde{t}_{\mathrm{p}}}
\end{equation}
\begin{equation}
\label{eqsolvtp2n}
\frac{\mathrm{d}^{3}\tilde{V}_{l}(\tilde{t}_{\mathrm{p}})}{\mathrm{d}\tilde{t}_{\mathrm{p}}^{\,3}}\left(\tilde{t}-\tilde{t}_{\mathrm{p}}\right) <2 \frac{\mathrm{d}^{2}\tilde{V}_{l}(\tilde{t}_{\mathrm{p}})}{\mathrm{d}\tilde{t}_{\mathrm{p}}^{\,2}}
\end{equation}
\begin{equation}
\label{eqsolvtp3n}
0\leq \tilde{t}-\tilde{t}_{\mathrm{p}} \leq \gamma\left/\frac{\mathrm{d}\tilde{V}_{l}(\tilde{t}_{\mathrm{p}})}{\mathrm{d}\tilde{t}_{\mathrm{p}}}\right..
\end{equation}

When the second inequality in formula (\ref{eqsolvtp3n}) becomes equality, one can determine
the moment when the shock just begins to happen, and therefore the solution of the 
shock parcel ejection time $\tilde{t}_{\mathrm{s}} = \tilde{t}_{\mathrm{p}}$ can be derived by solving 
formulas (\ref{eqsolvtp1n}, \ref{eqsolvtp2n}, \ref{eqsolvtp3n}) for this moment. In this case, 
the shock time can also be 
derived from the second equality of formula (\ref{eqsolvtp3n}) as
\begin{equation}
\label{eqtc}
\tilde{t}_{\mathrm{c}}=\tilde{t}_{\mathrm{s}}+\gamma\left/\frac{\mathrm{d}\tilde{V}_{l}(\tilde{t}_{\mathrm{s}})}{\mathrm{d}\tilde{t}_{\mathrm{s}}}\right..
\end{equation}
With $\tilde{t}_{\mathrm{p}}, \tilde{t}_{\mathrm{s}}, \tilde{t}_{\mathrm{c}}$ 
solved from the formulas, one can 
determine the density peak or shock region postions from
Eqs.~(\ref{eqrpeakn}) and (\ref{eqrshockn}),
and hence the pattern geometry.

Semi-analytical solutions for the special case of circular orbit have been 
derived on the basis of the formulas. (The solutions are not shown in this paper.)
Archimedes' spiral pattern 
and alternative concentric shell pattern have been confirmed by the analytical solutions 
only in the outer part of the pattern, while, in the central region of the patterns, 
deviation from these standard curves is expected.

\end{appendix}

\end{document}